\documentclass[a4paper,fleqn,usenatbib]{mnras}

\usepackage{mathptmx}

\usepackage[T1]{fontenc}
\usepackage{ae,aecompl}

\usepackage{graphicx}	
\usepackage{amsmath}	
\usepackage{amssymb}	

\title[Planetesimals in Rarefied Gas]{Planetesimals in Rarefied Gas: Wind Erosion in Slip Flow}

\author[T. Demirci et al.]{
	Tunahan Demirci$^{1}$\thanks{E-mail: tunahan.demirci@uni-due.de (TD)},
	Niclas Schneider$^{1}$,
	Tobias Steinpilz$^{1}$,
	Tabea Bogdan$^{1}$,\newauthor
	Jens Teiser$^{1}$
	and Gerhard Wurm$^{1}$\\
	$^{1}$	University of Duisburg-Essen, Faculty of Physics, Lotharstr. 1, 47057 Duisburg, Germany
}

\date{Accepted 2020 February 27; Received 2019 October 18; in original form ZZZ}

\pubyear{2020}

\begin{document}
	\label{firstpage}
	\pagerange{\pageref{firstpage}--\pageref{lastpage}}
	\maketitle
	
	\begin{abstract}
		A planetesimal moves through the gas of its protoplanetary disc where it experiences a head wind. Though the ambient pressure is low, this wind can erode and ultimately destroy the planetesimal if the flow is strong enough. For the first time, we observe wind erosion in ground based and microgravity experiments at pressures relevant in protoplanetary discs, i.e. down to $10^{-1}\, \rm mbar$.
		We find that the required shear stress for erosion depends on the Knudsen number related to the grains at the surface. The critical shear stress to initiate erosion increases as particles become comparable to or larger than the mean free path of the gas molecules. This makes pebble pile planetesimals more stable at lower pressure. However, it does not save them as the experiments also show that the critical shear stress to initiate erosion is very low for sub-millimetre sized grains.
	\end{abstract}
	
	
	
	
	\section{Introduction}
	\label{sec:introduction}
	
	There has been tremendeous improvement in our understanding of the first phases of planet formation over the last decades. On one side the idea that collisional growth rules from micrometre dust to millimetre size survived the years and can be considered as robust fact by now \citep{Blum2008, Johansen2014, Wurm2018}. Collisions mostly lead to bouncing then \citep{Zsom2010, Kruss2016, Demirci2017}. On the other side gravitational instability of a dense dust cloud once suggested by \citet{Goldreich1973} and \citet{Safronov1969} submerged for a while. However, it reappeared in recent years \citep{Johansen2006,Chiang2010,Dittrich2013}. Stitching these two mechanisms together can generate planetesimals which are consisting of very loosely bound mm dust granules not yet compacted by gravity \citep{Blum2017, Wahlberg2017}.
	
	Planetesimals born this way continue their lives moving within the gas of the surrounding protoplanetary disc. At minimum, the relative velocity between gas and planet on circular orbits is set by the pressure gradient of the disc and is typically on the order of $50\,\mathrm{m}\,\mathrm{s}^{-1}$  \citep{Weidenschilling1977}. This is only 1 per mille of the absolute orbital velocity. That means that even small changes in orbit can provide much larger relative velocities between planetesimal and gas. If the planetesimal orbit becomes only slightly non-circular, e.g. assuming Earth's eccentricity of only 0.016, the maximum flow velocities at perihelion  -- though still small compared to the absolute speed -- now already reach a factor of ten more or $500\,\mathrm{m}\,\mathrm{s}^{-1}$. As, generally speaking, the dynamic pressure is proportional to the square of the flow velocity, this might be significant. 
	
	Even if the gas density is low in protoplanetary discs, gravity on a planetesimal is also low. It is therefore likely that conditions are met somewhere in protoplanetary discs under which a planetesimal with almost no cohesion will lose mass to the protoplanetary wind engulfing it, noting that we really mean wind here, not particle radiation.
	
	First estimates from laboratory wind tunnel experiments at millibar pressure by \citet{Paraskov2006} confirmed that the flow is indeed erosive on only slightly eccentric orbits in the inner part of protoplanetary discs. These experiments did not have pebble pile planetesimals in mind, as these were not really predicted yet. They used dust with significant cohesion and found that a shear stress of $5\,\mathrm{Pa}$ was necessary to erode a dusty surface under Earth gravity. In this case, shear stress has to overcome gravity as dominating force though. The basic results on the required shear stress were in agreement to \citet{Greeley1980}.  
	
	\citet{White1987} studied the influence of gravity on erosion and found that the threshold shear stress is linearly dependent on the gravitational acceleration. These experiments were carried out at normal pressure. In recent microgravity experiments \cite{Musiolik2018} and \citet{Kruss2019} studied the threshold conditions for erosion at gravity levels lower than on Earth and at low pressure conditions , especially of Mars ($6\,\mathrm{mbar}$). 
	The latest experiment of that type pushed the gravity level and ambient pressure to the lowest levels so far and was carried out by \citet{Demirci2019}. Using $425\,\mu\mathrm{m}$ (diameter) grains and at pressures down to a level of a few mbar, they found that the gravity dependence of shear stress still follows existing models, i.e. by \citet{Shao2000} (see below for details).
	
	\citet{Demirci2019} then extrapolated to planetesimals in protoplanetary discs and showed that depending on grain size, surface energy (sticking properties of grains) and gas density planetesimals are not stable inside a certain distance from the star.
	
	In any case, the experiments carried out so far were still not in the range of pressures found in protoplanetary discs. Again, carrying out such experiments at very low pressure requires microgravity to lower the necessary threshold shear stress.
	We therefore designed and built a new experiment which we carried out for the first time on a parabolic flight campaign in early 2019. We complemented this study by ground based experiments with the same setup. We report on these experiments here.

	\section{Wind Erosion in Slip Flow}
	\label{sec:winderosion}
	
	In the Minimum Mass Solar Nebula the mean free path of the gas molecules at $1\, \rm AU$ is a few centimetre \citep{Hayashi1981}. This is small with respect to a planetesimal but is comparable in size with respect to the pebbles which are the building blocks of the planetesimal.
	Therefore, with respect to the gas flow, there are two size scales important for eolian erosion.
	
	The general flow around the planetesimal might be considered as hydrodynamic. From that point of view, the gas does not move at the surface but only increases in speed $u$ with increasing height $h$ above the surface.
	This view finds itself in the definition of the dynamic viscosity $\eta$ or \citep{Schlichting2006}
	\begin{equation}
		\label{eq:taudefinition}
		\tau = \eta \left. \frac{\partial u}{\partial h}\right|_{h=0},
	\end{equation}
	where $\tau$ is the shear stress at the surface $(h=0)$. The shear stress can also be used to define the friction velocity $u^*$ at the given gas density $\rho$ as
	\begin{equation}
		\label{eq:frictionveldefinition}
		\rho u^{*2} := \tau .
	\end{equation}
	At small heights the velocity gradient can always be considered to be constant (linear increase with height) \citep{Merrison2008,Musiolik2018,Demirci2019}. As our flow is laminar and generated with a second surface rotating at a certain speed this also holds for all heights in the given experiment (see below). In any case, then $\partial u/\partial h = u/h$ and the gas velocity can be expressed as
	\begin{equation}
		\label{eq:gasvelocity}
		u(h) = \frac{\tau}{\eta}h.
	\end{equation}
	
	This picture only changes if the flow around an object changes from hydrodynamic to free molecular. 
	In this case, the gas velocity on the surface of an object can no longer be considered to be zero. As molecules are now reflected and only hit other molecules much later, the gas literally "slips" over the body's surface.
	Slip flow gets important if the mean free path of the gas molecules $\lambda$ becomes comparable to the size $d$ of the object it flows around. The ratio of this two quantities is the Knudsen number
	\begin{equation}
		\mathrm{Kn}=\frac{\lambda}{d}=\frac{k_\mathrm{B}}{\sqrt{2} \pi d_\mathrm{mol}^2 R_\mathrm{s} \rho} \frac{1}{d},
	\end{equation}
	with $k_\mathrm{B}$ being is the Boltzmann constant, $d_\mathrm{mol}$ the molecule diameter and $R_\mathrm{s}$ the specific gas constant. 
	
	For a kilometre-size planetesimal, the Knudsen number is small and all flows around planetesimals important here are hydrodynamic in nature. However, on the size scale of a constituent pebble, the flow can become molecular. Therefore, slip flow becomes important for individual grains on the surface of the planetesimal.
	
	With the onset of slip flow the drag force on an object is increasingly reduced. The concept of hydrodynamic drag can still be used but has to be modified by a correction factor, e.g. by the Cunningham factor \citep{Davies1945}
	\begin{equation}
		f_\mathrm{C}(\mathrm{Kn})=1+2 \mathrm{Kn} \left[1.257 + 0.4 \exp\left(- \frac{0.55}{\mathrm{Kn}}  \right)  \right].
	\end{equation}
	This correction is applicable for Knudsen numbers up to a few hundred. The gas drag force on an object is then given as
	\begin{equation}
		F_\mathrm{drag}=\frac{1}{2}\rho u^2 A C_\mathrm{D} \frac{1}{f_\mathrm{C}(\mathrm{Kn})}.
	\end{equation}
	Here, $u$ is the relative velocity between object and gas, $A=\pi {d^2}/{4}$ is the cross-section of the object and $C_\mathrm{D}$ its dimensionless drag coefficient. The drag coefficient depends on the Reynolds number
	\begin{equation}
		\mathrm{Re} = \frac{\rho u d}{\eta}.
	\end{equation}
	For small Reynolds numbers $\mathrm{Re}<1$ the drag coefficient of a sphere is approximately $C_\mathrm{D}\approx 24/\mathrm{Re}$. This results in the corrected Stokes drag force for pebbles on planetesimal surfaces
	\begin{equation}
		\label{eq:drag_stokes}
		F_\mathrm{drag} = 3 \pi \eta u d \frac{1}{f_\mathrm{C}(\mathrm{Kn})}.
	\end{equation}
	If we now assume -- on planetesimal size -- a hydrodynamic flow at the surface (Eq. \ref{eq:gasvelocity}) but a slip flow around a pebble on top of this surface, the drag force on the pebble, using Eq. \ref{eq:drag_stokes}, is
	\begin{equation}
		\label{eq:drag_stokes_tau}
		F_\mathrm{drag}=\frac{3}{2} \pi \tau d^2 \frac{1}{f_\mathrm{C}(\mathrm{Kn})},
	\end{equation}
	if we take the velocity at half the particle height.
	This way, the gas flows on both scales (hydrodynamic flow and slip flow) set the drag force acting on the particle.
	
	In an idealized view, this drag force acts along the surface and it is the grain's friction force which this drag has to overcome to initiate particle motion and, ultimately, particle detachment. The friction force is perpendicular to the pebble's gravitational and cohesive force with an effective friction coefficient $\alpha$. Therefore, we get for the threshold condition
	\begin{equation}
		\label{eq:force_balance}
		F_\mathrm{drag} = \alpha \left( F_\mathrm{gravity}+F_\mathrm{cohesion} \right),
	\end{equation}
	The pebble's gravitational force is
	\begin{equation}
		F_\mathrm{gravity}=\frac{1}{6}\pi \rho_\mathrm{p} g d^3
	\end{equation}
	with the gravitational acceleration $g$ and the particle mass density $\rho_p$. The cohesion force between the pebbles is
	\begin{equation}
		\label{eq:jkr}
		F_\mathrm{cohesion}=\frac{3}{2}\pi \gamma {d},
	\end{equation}
	where $\gamma$ is the surface energy \citep{Johnson1971}. 
	Neither of these two can be neglected for planetesimals \textit{a priori} with gravity and cohesion both being small.
	
	Note that the friction force is significantly lower than the force needed to pull off a grain by compensating gravity and cohesion completely or, in other words, $\alpha$ is small. Following \citet{Zimon1982} we can estimate $\alpha = \frac{r_\mathrm{c}}{2d}$, with $r_\mathrm{c}$ being the contact radius. Using experimental values, this yields $\alpha$ on the order of $10^{-3}$, but we keep $\alpha$ as fit parameter for a real particle bed.

	Finally, resolving the force balance for the threshold of particle detachment for $\tau_\mathrm{erosion}$ yields the threshold shear stress, which is needed to erode a pebble planetesimal
	\begin{equation}
		\label{eq:tau_erosion}
		\tau_\mathrm{erosion}=\alpha f_\mathrm{C}(\mathrm{Kn}) \left(\frac{1}{9}\rho_\mathrm{p} g d +\frac{\gamma}{d}  \right).
	\end{equation}
	The expression for the threshold friction velocity is then (Eq. \ref{eq:taudefinition})
	\begin{equation}
		u^*_\mathrm{erosion}=\sqrt{\alpha f_\mathrm{C}(\mathrm{Kn}) \left(\frac{1}{9}\frac{\rho_\mathrm{p}}{ \rho} g d + \frac{\gamma}{\rho d}  \right)}
	\end{equation}
	This equation has the same dependencies as in the erosion model by \citet{Shao2000}. A significant difference though is the Cunningham correction $f_\mathrm{C}(\mathrm{Kn})$ now, which scales the threshold shear stress or threshold friction velocity in the slip flow regime. In comparison to the erosion model by \citet{Shao2000}, $\alpha \propto A_\mathrm{N}^2$ gets a similar interpretation as the dimensionless threshold friction velocity $A_\mathrm{N}$.
	
	This derivation combines elements of hydrodynamic flow and molecular flow, which might not describe the drag force accurately. We therefore carried out the experiments to test this symbiotic approach.
	
	\section{Experimental setup \& procedures}
	\label{sec:microgravityexperiments}
	
	\begin{figure*}
		\centering
		\includegraphics[width=\textwidth]{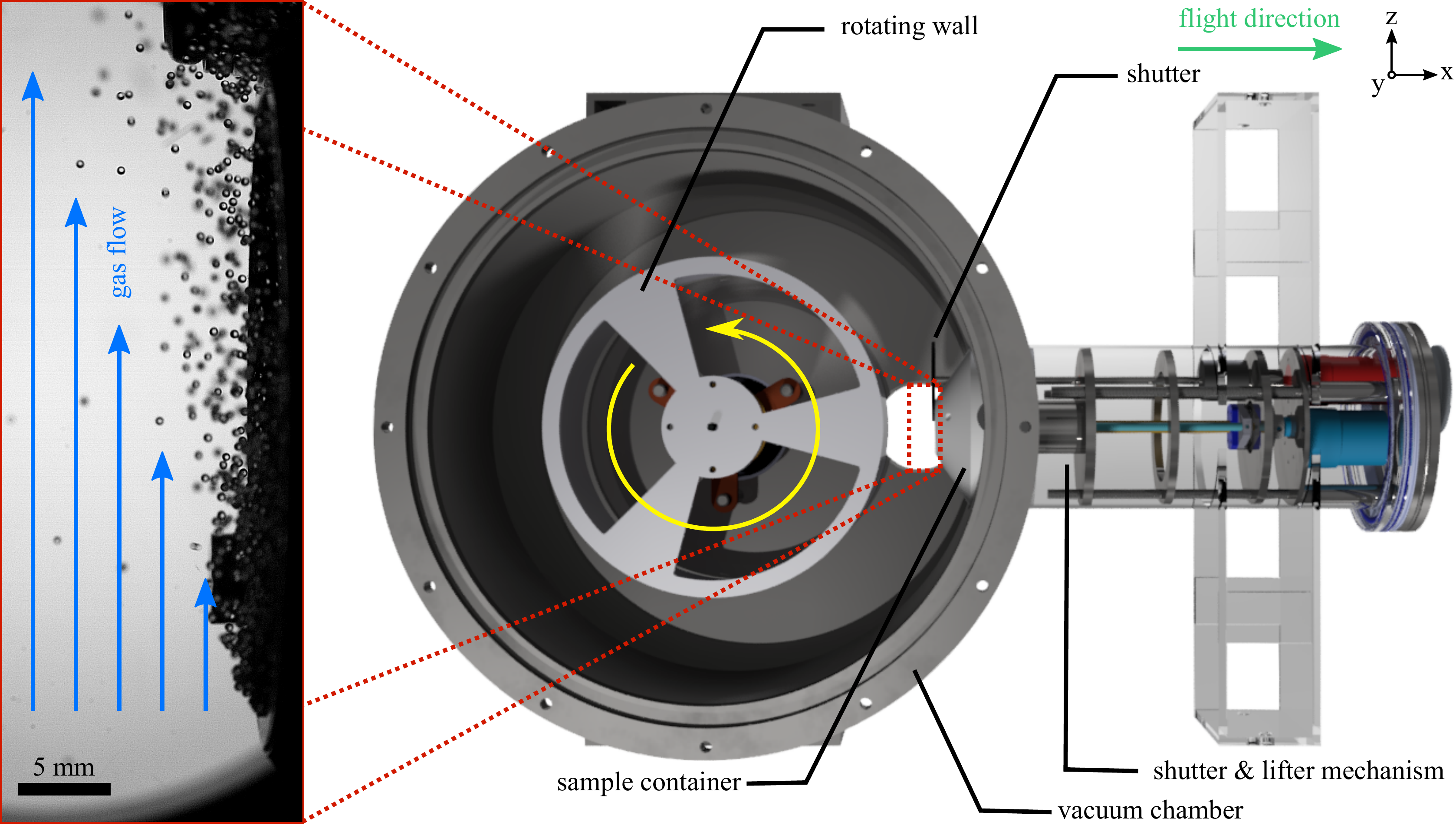}
		\caption{\label{fig:setup} Experimental setup: A rotating wall inside a vacuum chamber creates a shear flow with a maximum velocity of up to $125\,\mathrm{m}\,\mathrm{s}^{-1}$ ($\equiv 200\,\mathrm{Hz}$ rotation frequency). With this mechanism a linear gas flow profile can be generated at low ambient pressures down to $10^{-2}\,\mathrm{mbar}$. This setup is mainly used for parabolic flight experiments. Therefore, the sample container is placed in flight direction, because the residual acceleration $g \approx 0.01 g_\mathrm{earth}$ during a parabola acts in this direction. The shutter prevents the sample from spilling out of the sample container and opens only, if the residual acceleration is strong enough in this direction. The sample bed is observed in perpendicular direction to the wind flow with a high-speed camera with a frequency of $939\,\mathrm{Hz}$. A snapshot of the eroding sample bed is shown on the left side of this figure.}
	\end{figure*}
	\subsection{Experimental setup}
	
	\begin{figure}
		\centering
		\includegraphics[width=\columnwidth]{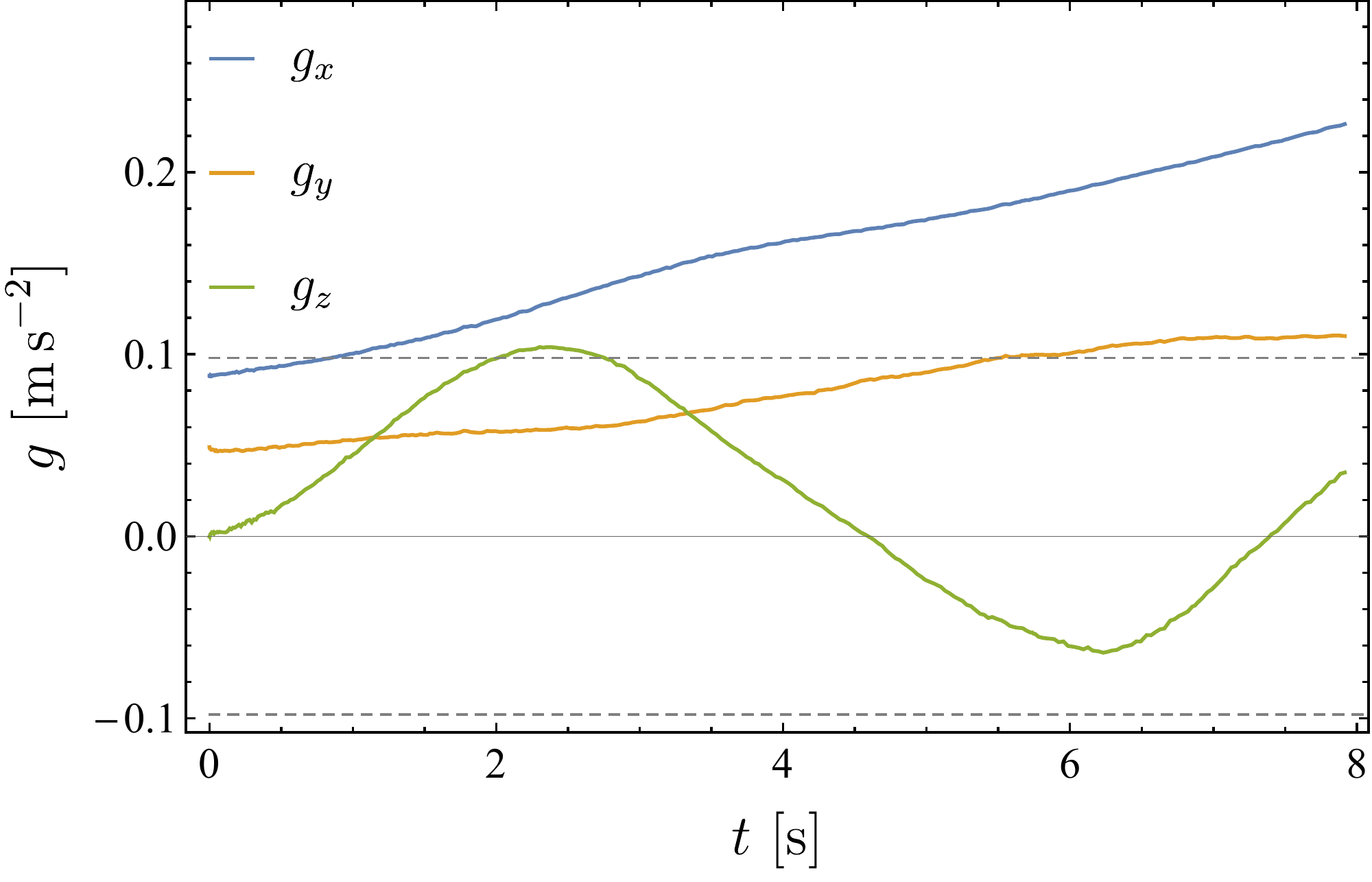}
		\caption{\label{fig.residual_acceleration} Residual acceleration during an example parabola. The shutter to expose the sample is open only, if $g_x>0.01\,g_\mathrm{earth}$ and $|g_y|,|g_z|<0.01\,g_\mathrm{earth}$.}
	\end{figure}
	
	Fig. \ref{fig:setup} shows a sketch of the setup of the experiment. 
	In principle, a rotating wheel with a diameter of $20\,\mathrm{cm}$ inside a cylinder (vacuum chamber with a diameter of $32\,\mathrm{cm}$) generates a shear flow at pre-determined ambient pressure. The maximum rotation frequency is $200\,\mathrm{Hz}$ which corresponds to a tangential velocity of $125\,\mathrm{m}\,\mathrm{s}^{-1}$ of the rotating wall. The rotating wheel is made of aluminum and weighs $0.7\, \rm kg$. The motor for the rotating wheel is a brushless motor (\textit{Maxon EC-4pole 468313}) with three Hall sensors measuring the rotation frequency.
	
	The vacuum chamber can be evacuated to an ambient pressure down to $10^{-4}\,\mathrm{mbar}$ with a membrane pump and a turbo-molecular pump. With this setup, a linear gas flow profile can be created at low pressures down to $10^{-2}\,\mathrm{mbar}$. At lower gas densities the general gas flow becomes molecular and the shear flow does not work well. The wind profile in detail and its determination with the particle tracking velocimetry technique is described in sec. \ref{sec:wind_profile}.
	
	The experiment is designed to be mainly used for parabolic flight experiments. The parabolic flights are performed with the \textit{Airbus A310 ZERO-G} by Novespace in Bordeaux, France \citep{Pletser2016}. The sample bed is oriented in the plane in flight direction to take advantage of the residual acceleration in $x$-direction during the weightlessness phase of a parabola. The residual acceleration (g-jitter) in this direction is systematic in contrast to the other directions. This is due to the rotation of the aircraft along the $y$-axis during the maneuver. Starting at $-0.02g_\mathrm{earth}$ it linearly increases to $0.02g_\mathrm{earth}$ during a parabola. Here, $g_{\mathrm{earth}}=9.81\, \mathrm{m}\,\mathrm{s}^{-2}$ is the gravitational acceleration on Earth's surface. The residual accelerations in all three directions in the relevant time range are shown in Fig. \ref{fig.residual_acceleration}. The g-jitter in $y$- and $z$-direction are not systematic and are kept as small as possible by the pilots. If the residual acceleration points in flight direction, the sample will be exposed to the wind by opening the shutter. The shutter is only open, while $g_x>0.01\,g_\mathrm{earth}$ and $|g_y|,|g_z|<0.01\,g_\mathrm{earth}$. We measure the accelerations on our experiment rack with a triaxial MEMS DC accelerometer (\textit{PCB Piezotronics 3713B112G}), which has a relative error of $\pm 5\,\%$ for accelerations between $\pm 2\,g_\mathrm{earth}$ and frequencies up to $250\,\rm Hz$. If the conditions are no longer met, the shutter closes automatically. The shutter prevents the sample from spilling out of the sample container. A motor driven sample lift ensures that the sample is lifted up to expose the surface to the wind at the beginning of each experiment. In ground based experiments the setup was turned by 90 degrees so that the sample bed is oriented to the bottom.
	
	The surface and onset of erosion of a particle bed in this wind are observed by a high-speed camera with a frequency of $939\,\mathrm{Hz}$. The sample bed is observed in perpendicular direction to the wind flow. A snapshot of the eroding sample bed is shown on the left side of Fig. \ref{fig:setup}.
	
	The vibrations of the rotating wheel and of the aircraft also cause accelerations which are on the order of $a_\mathrm{vib}=0.002\,g_\mathrm{earth}$. We determined the acceleration due to the vibrations by evaluating the standard deviation of the g-sensor values. The acceleration is small compared to the g-jitter, but increases from a rotation frequency of $120\,\rm Hz$ to higher values ($0.01\,g_\mathrm{earth}$ at $150\,\rm Hz$ and $0.035\,g_\mathrm{earth}$ at $200\,\rm Hz$). All experimental runs, presented in this work, are below $f=120\,\rm Hz$ ($a_\mathrm{vib}\leq 0.002\,g_\mathrm{earth}$).
	
	On the recent parabolic flight campaign (33rd DLR campaign, 4th-15th March 2019) we used spherical glass beads with a diameter $d=425\,\mu \mathrm{m}$ as sample. The particle mass density is $\rho_\mathrm{p}=2460 \pm 30\,\mathrm{kg}\,\mathrm{m}^{-3}$. The reason for choosing this sample is that they behave like pebbles with low cohesion on pebble pile planetesimals. This sample was also used by \citet{Demirci2019}, so that the results can be compared with their data.
	
	For the first time we tested the performance of this new experimental setup on the parabolic flight campaign. For ambient pressures between $10^{-3}$ and $10^{-1}\,\mathrm{mbar}$ wind erosion experiments were carried out. The threshold shear stress for wind erosion in dependence of the pressure was determined by probing different frequencies of the rotating wall whether erosion occurs or not. Before each parabola the desired rotation frequency and ambient pressure was adjusted in discrete steps. We also performed ground based experiments on wind erosion at $1\,g_\mathrm{earth}$. For these measurements we used two different sizes of spherical glass beads ($425\,\mu\mathrm{m}$ and $90\,\mu\mathrm{m}$). 
	
	\subsection{Trajectory analysis \& wind profile}
	\label{sec:wind_profile}
	
	\begin{figure}
		\centering
		\includegraphics[width=\columnwidth]{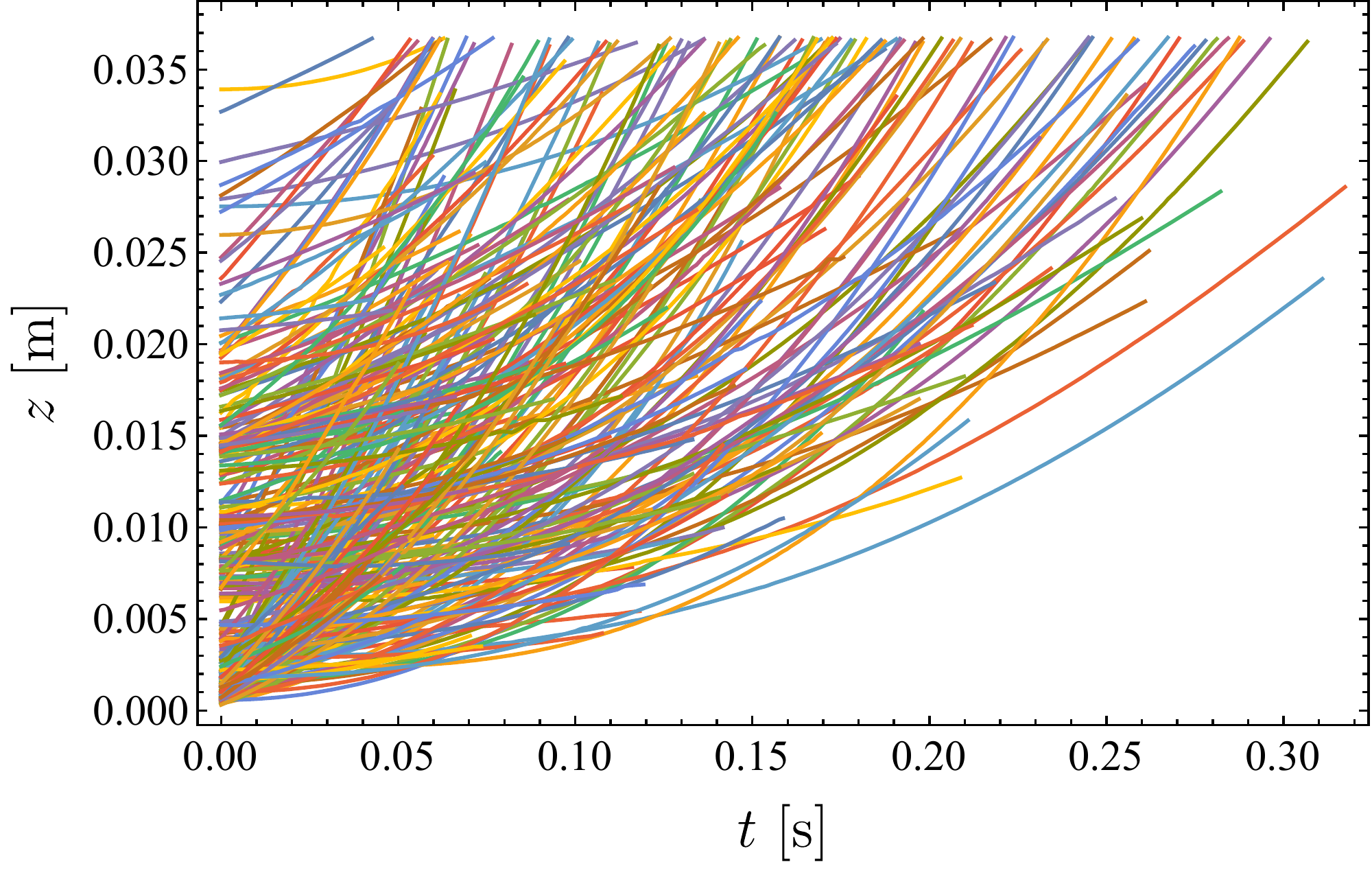}
		\caption{\label{fig.z(t,h)} Evaluated trajectories $z(t,h)$ for the experimental run at $0.018\,g_\mathrm{earth}$ and $10^{-1}\,\mathrm{mbar}$. The gas flow is determined by fitting equation \ref{eq:acceleration} to the trajectories. Only fits with a high adjusted coefficient of determination ($\overline{R}^2>0.999$) are considered for the wind profile determination.}
	\end{figure}
	
	\begin{figure}
		\centering
		\includegraphics[width=\columnwidth]{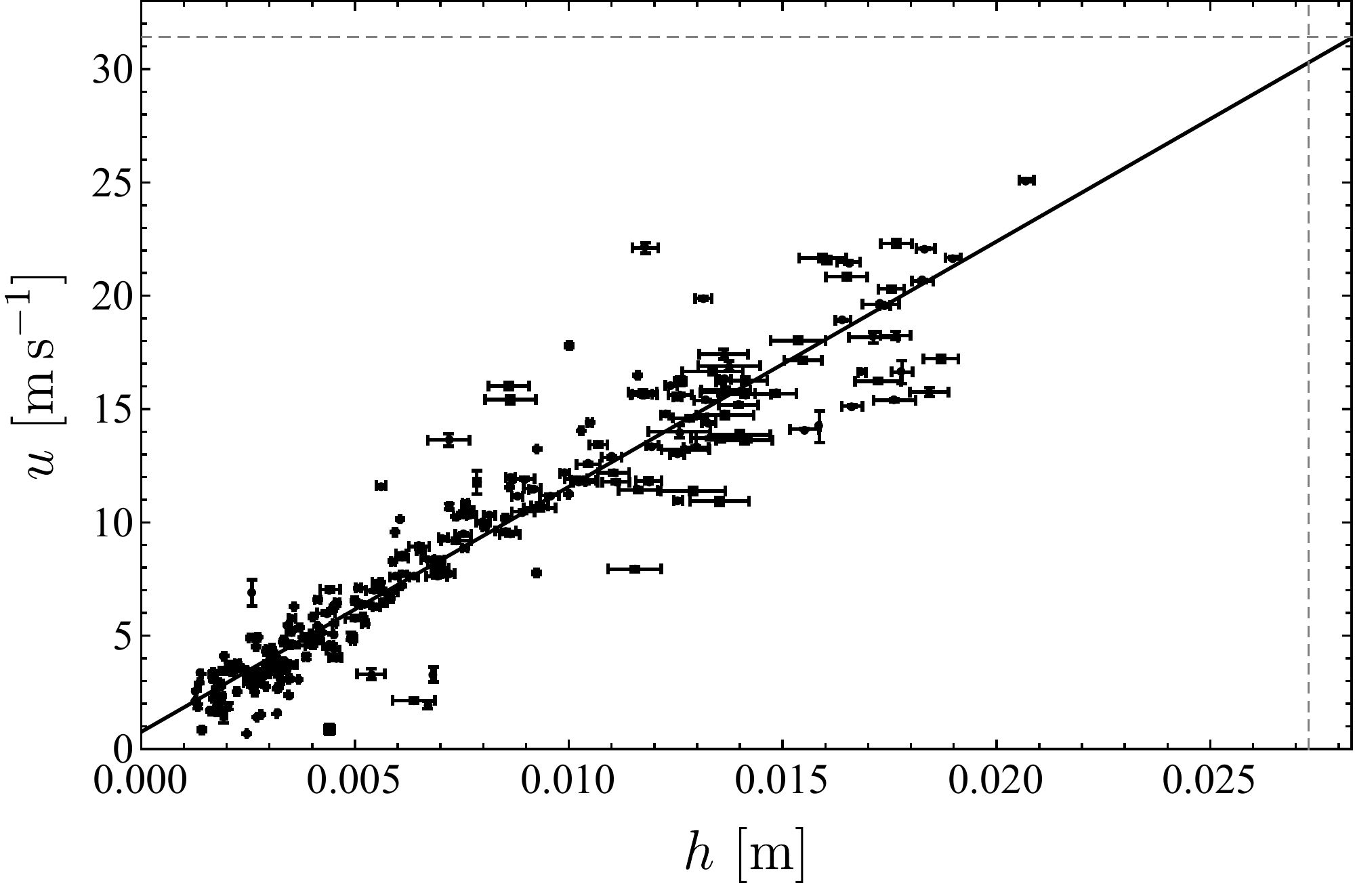}
		\caption{\label{fig.u(h)} Gas flow profile $u(h)$ at the threshold shear stress for wind erosion at $0.018\,g$ and $10^{-1}\,\mathrm{mbar}$. The wall is rotating with a velocity of $31.4\,\mathrm{m}\,\mathrm{s}^{-1}$ at a height of $h=0.0273\,\mathrm{m}$ (indicated as gray dashed lines).}
	\end{figure}
	
	For the determination of the shear stress applied on the surface of the sample bed we evaluated the gas flow profile. The gas flow velocity in dependence of the height can be determined by the analysis of the trajectories of lifted beads. This technique is called particle tracking velocimetry and is often used for gas flow characterizations at small ambient pressures \citep{Musiolik2018,Demirci2019,Kruss2019}. We use the \textit{Fiji} plug-in \textit{TrackMate} as tracking tool \citep{Schindelin2012,Tinevez2017} like \citet{Demirci2019} and \citet{Schneider2019}. Sample spheres, which are lifted by the wind, couple to the gas and are accelerated in flow direction within the coupling time $t_\mathrm{c}$. The motion in flow direction of the lifted spheres is described by the following equation \citep{Wurm2001}
	\begin{equation}
		\label{eq:z(t,h)}
		z(t,h)=z_0+ \left( u(h)-v_0 \right) t_\mathrm{c}\exp\left(-\frac{t}{t_\mathrm{c}} \right) +u(h) t,
	\end{equation}
	with $z_0$, $u(h)$, $v_0$ and $t_\mathrm{c}$ being constants. The coupling time for a sphere for Stokes-drag ($\mathrm{Re}<1$) and slip flow (Cunningham correction) is
	\begin{equation}
		t_\mathrm{c}=\frac{m u}{F_\mathrm{Drag}}=\frac{1}{18} f_\mathrm{C}(\mathrm{Kn})\frac{\rho_\mathrm{p} d^2}{\eta}.
	\end{equation}
	For a sphere with $d=90\,\mu \mathrm{m}$ at an ambient pressure $p=10\,\mathrm{mbar}$ it is $t_\mathrm{c}=0.07\,\mathrm{s}$ and the duration of the trajectories of the lifted beads are longer than the coupling time. Then, Eq. \ref{eq:z(t,h)} can be fitted to the trajectories. The fit of each trajectory will yield a gas flow velocity $u(h)$ at a certain height $h$. For a sphere with $d=425\,\mu \mathrm{m}$ at $p=10^{-1}\,\mathrm{mbar}$ the coupling time is $t_\mathrm{c}=8.14\,\mathrm{s}$ and thus much longer than the duration of the trajectories. Fitting Eq. \ref{eq:z(t,h)} does not work well, because the exponential decay is not significant during the tracks. Nevertheless, the spheres are accelerated by the wind. We analyze the beginning of the acceleration (approx. $0.1-0.3\,\mathrm{s}$, see Fig. \ref{fig.z(t,h)}) by considering the Taylor series of degree $2$ of Eq. \ref{eq:z(t,h)}. Then, for $t\rightarrow 0$ $z(t,h)$ can be approximated with
	\begin{equation}
		\label{eq:acceleration}
		z(t,h) \approx z_0+(u(h)-v_0)t_\mathrm{c} + v_0 t +\frac{1}{2}\frac{1}{t_\mathrm{c}}(u(h)-v_0)t^2.
	\end{equation}
	Here, $u(h)-v_0$ is the initial speed of the sphere and $a=\frac{1}{t_\mathrm{c}}(u(h)-v_0)$ its acceleration. This acceleration does not contain the residual gravitational acceleration, but in reality the spheres are also accelerated due to the g-jitter. To obtain the gas flow velocity from the fit we need to correct the determined acceleration by subtracting the g-jitter in $z$-direction at the moment $t_i$ of each track $i$ 
	\begin{equation}
		u(h)=\left(a-g_z(t_i)\right)t_\mathrm{c}+v_0.
	\end{equation}
	This kind of g-jitter correction is usually done by evaluating parabolic flight experiments \citep{Steinpilz2017,Kraemer2019}. For ambient pressures $p>10^{-2}\,\mathrm{mbar}$ the acceleration due to the gas drag is larger than the residual acceleration in $z$-direction due to the g-jitter (see Fig. \ref{fig.residual_acceleration}). Fig. \ref{fig.z(t,h)} shows trajectories for the experimental run at $0.018\,g_\mathrm{earth}$ and $0.1\,\mathrm{mbar}$. The gas flow is determined by fitting equation \ref{eq:acceleration} to the trajectories. Only fits with a high adjusted coefficient of determination ($\overline{R}^2>0.999$) are considered for the wind profile determination.
	
	Using the trajectories to quantify the gas flow velocities results in linear velocity profiles with height. An example for wind speeds measured for an experiment with a rotating surface with $31.4\,\mathrm{m}\,\mathrm{s}^{-1}$ is seen in Fig. \ref{fig.u(h)}. Extrapolated to the distance of the rotating surface, the evaluated velocities are consistent with a linear profile throughout the whole wind tunnel. This has to be expected as the Reynolds number of the tunnel is $\mathrm{Re} \approx 5$ and the flow is laminar.
	
	\section{Experimental Results}
	
	\subsection{Experiments under Earth gravity}
	
\begin{figure}
		\centering
		\includegraphics[width=\columnwidth]{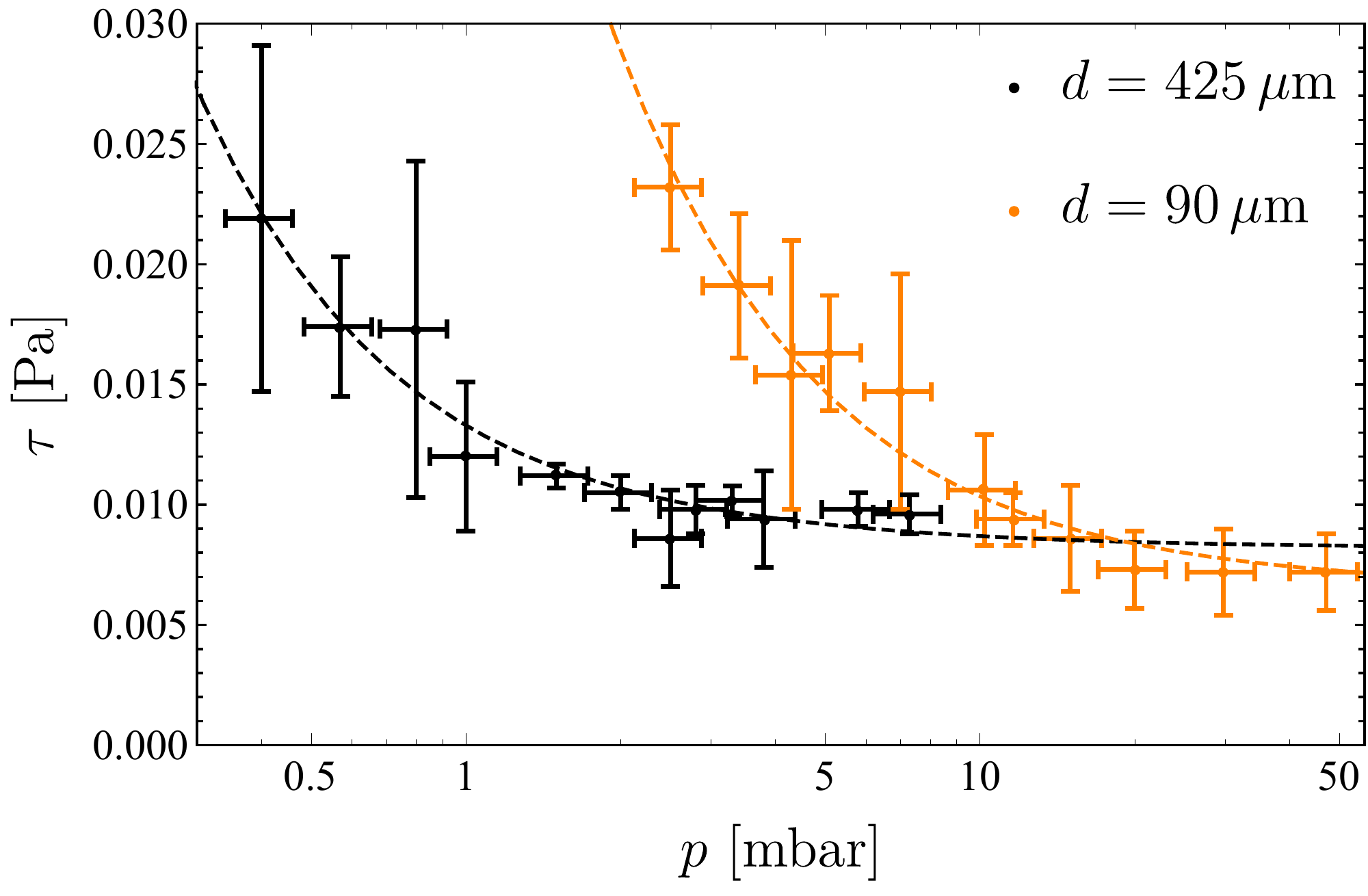}
		\caption{\label{fig.shear_stress(pressure)} Shear stress $\tau(p)$ in dependence of the pressure for spherical glass beads with $d=425\,\mu\mathrm{m}$ and $d=90\,\mu\mathrm{m}$ at Earth gravity.}
	\end{figure}

\begin{figure}
		\centering
		\includegraphics[width=\columnwidth]{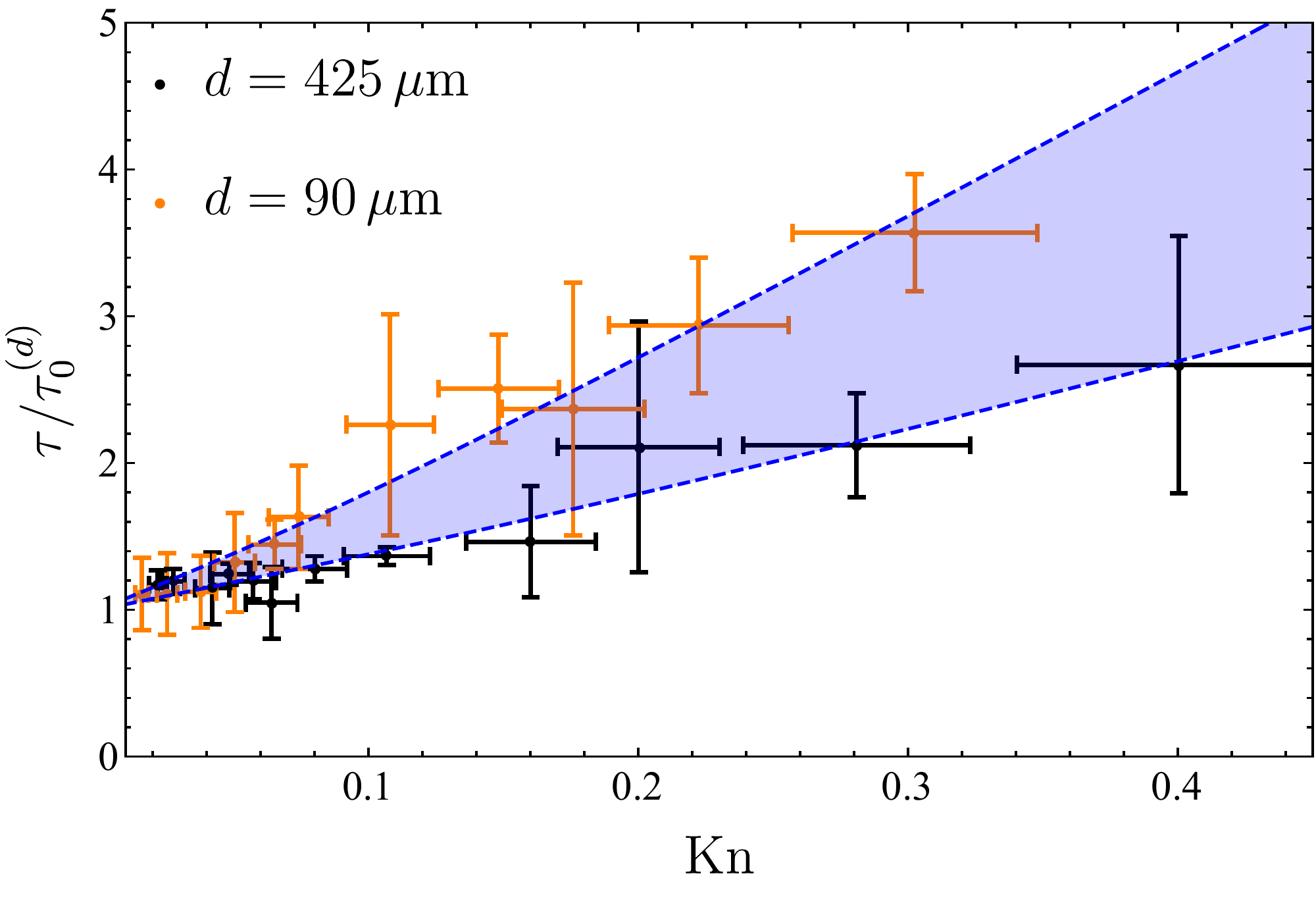}
		\caption{\label{fig.shear_stress(Knudsen)} Shear stress $\tau(p)$ normalized by the convergence value $\tau_0^{(d)}$ (see Fig. \ref{fig.shear_stress(pressure)}) in dependence of the Knudsen number for spherical glass beads with $d=425\,\mu\mathrm{m}$ and $d=90\,\mu\mathrm{m}$ at Earth gravity. The blue area marks the fit range for $d_\mathrm{eff}/d$ between $0.33$ and $0.67$.}
	\end{figure}

	Due to the high rotation speed and the close spacing between rotating inner wall and fixed outer wall, the new setup already provides a range in shear stress that allows erosion experiments at mbar pressure under Earth gravity. Fig. \ref{fig.shear_stress(pressure)} shows the pressure dependence of the threshold shear stress for two particle sizes.
	Both show the dependencies derived in Eq. \ref{eq:tau_erosion} and the dotted lines are fits to this equation, fitting $\alpha$, $\gamma$, and the size of the grains entering in $f_C$. The latter is left as free parameter as we cannot be sure that the simple argument given above is true. In any case, both datasets show the principle trends that at high pressure the drag and therefore the threshold shear stress is independent of the particle size and pressure. However, once the pressure is reached where the mean free path gets comparable to the grain size, the needed shear stress increases toward lower pressure. For the smaller particle size this turnover occurs already at higher pressure. This is again visualized in Fig. \ref{fig.shear_stress(Knudsen)}, which shows the same data but plotted over the Knudsen number. Also, the shear stress is normalized by the equilibrium shear stress at high pressure or small Knudsen numbers.
	With this normalization the increase with Knudsen number should be independent of the grain size, if $f_C$ only depends on the grain's Knudsen number. In detail, this is not the case. However, as mentioned before, we leave the grain size entering in the Knudsen number in $f_C$ as free parameter. This kind of consideration gives an effective Knudsen number of the surface grains. 
	
	For both sets the data can be fitted with one effective grain size that is a factor of 0.33 for the $90\,\mu \mathrm{m}$ grains and $0.67$ for the $425\,\mu \mathrm{m}$ grains or $d_\mathrm{eff} = \beta d$ with $\beta$ being 0.33 or 0.67, respectively. A correction factor $\beta$ has to be expected as, again, grains are not free floating but are part of a granular bed. Typically, grains are not fully exposed to the wind, but shielded by surrounding grains.
	
	Combining the analytic derivation and the laboratory experiments, the ultimate equation for the threshold shear stress is
	\begin{equation}
		\label{eq:tau_erosion_final}
		\tau_\mathrm{erosion}=\alpha f_\mathrm{C}(\beta^{-1} \mathrm{Kn}) \left(\frac{1}{9}\rho_\mathrm{p} g d +\frac{\gamma_\mathrm{eff}}{d}  \right).
	\end{equation}
	with the effective Knudsen number $\beta^{-1} \mathrm{Kn}$. There might be a size dependence but this has to be subject to future work as we could not measure a wider grain size range. As we expect pebbles with a size of $d=1\, \rm mm$ on pebble pile planetesimals, the value for the bigger probed glass spheres are more relevant and we currently suggest to use a constant $\beta = 0.67$.
	
	Depending on the sample we used (glass spheres) we get 
	$\alpha = (6.4 \pm 0.7) \times 10^{-3}$. As effective surface energy we find $\gamma_\mathrm{eff}=(7.3 \pm 3.8) \times 10^{-5} \,\mathrm{N}\,\mathrm{m}^{-1}$. We note, that $\gamma_\mathrm{eff}$ is many orders smaller than the surface energy for glass ($0.065\,\mathrm{N}\,\mathrm{m}^{-1}$, \citet{Mader-Arndt2014}). The reason for this low effective value is, that Eq. \ref{eq:jkr} is only applicable for perfectly smooth spheres. The glass spheres used in this work have asperities in the order of $d_\mathrm{asp}=0.3-0.7\,\mu \mathrm{m}$ \citep{Steinpilz2020}, which is similar to the contact radius of the spheres, so that the contacts between the pebbles are mainly determined by the asperities. To obtain the surface energy for the bulk material, instead of the pebble diameter $d$, the size of the asperities should be used in Eq. \ref{eq:jkr}. This leads to a surface energy $\gamma=\gamma_\mathrm{eff} \frac{d}{d_\mathrm{asp}}=0.04-0.1\, \mathrm{N}\,\mathrm{m}^{-1}$, which indeed is in agreement to the value of \citet{Mader-Arndt2014}. It is also similar to the surface energies of silicates expected in protoplanetary discs ($0.01 - 0.2\, \mathrm{N}\,\mathrm{m}^{-1}$, \citet{Heim1999, Tarasevich2006, Steinpilz2019, Kimura2015}). Therefore, the experiments simulate aggregates of silicate dust with few contacts as the formation process of pebble pile planetesimals might provide. Nevertheless, experiments with dust agglomerates, which are more realistic as protoplanetary pebble analogs, are planned in future works.

	\subsection{Erosion thresholds at reduced gravity}
	
\begin{figure}
		\centering
		\includegraphics[width=\columnwidth]{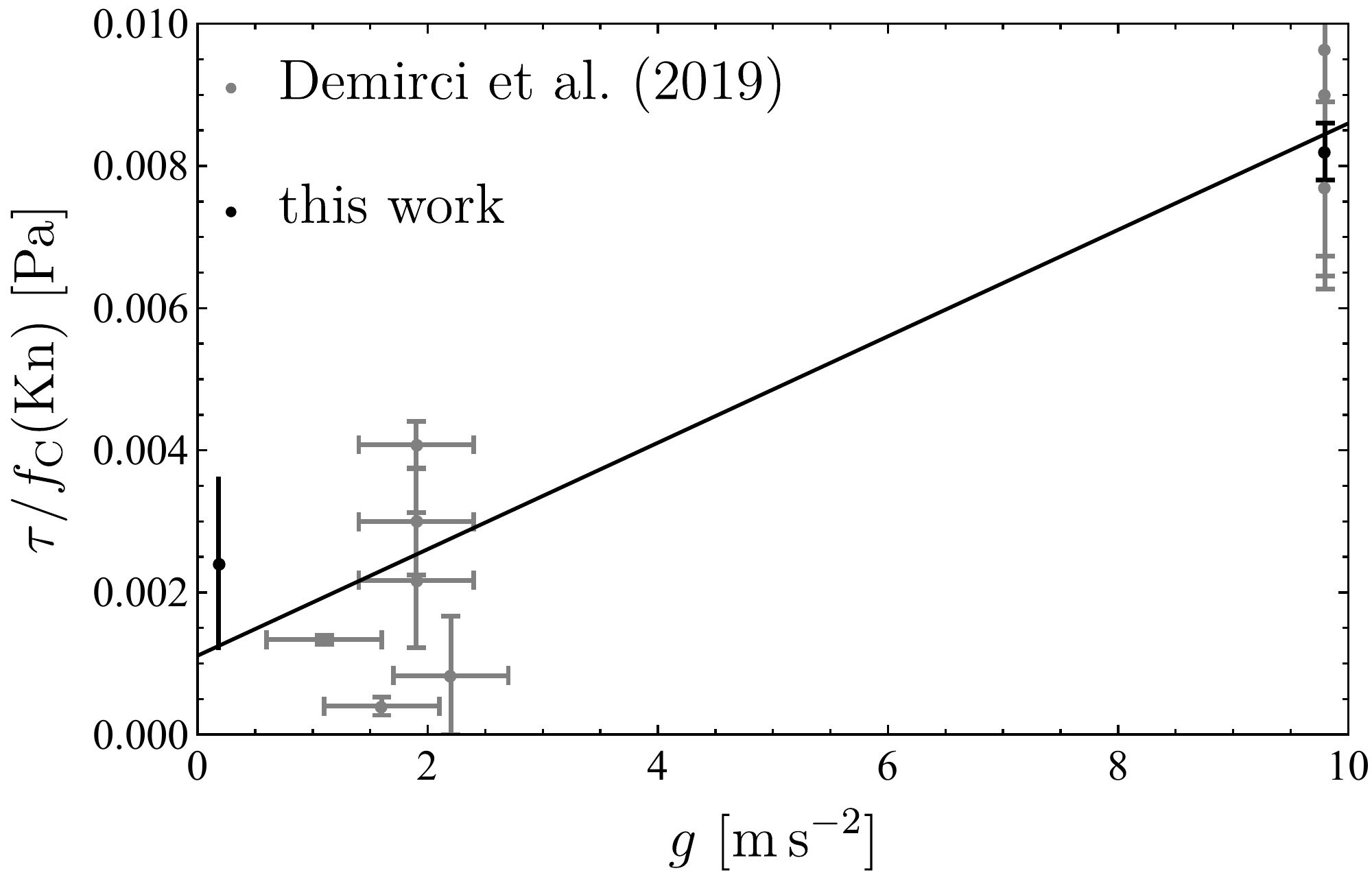}
		\caption{\label{fig.shear_stress(gravity)} Shear stress $\tau(g)$ divided by the Cunningham function $f_\mathrm{C}(\mathrm{Kn})$ in dependence of the gravitational acceleration for spherical glass beads with $d=425\,\mu\mathrm{m}$.}
	\end{figure}

	At Earth gravity, the shear stress always has to be strong enough to
	compensate the particles weight. At lower gravity, the surface energy part in Eq. \ref{eq:tau_erosion_final} gets important. At the same time, the ambient pressure can be reduced further to test still larger Knudsen numbers. 
	As the number of experiments on a parabolic flight is limited and as this was the first use of the new setup, only one firm erosion threshold for a gravity of $0.018\,g$ and a pressure of $10^{-1}\,\mathrm{mbar}$ was determined for the grain size of $425\,\mu\mathrm{m}$. The shear stress for these parameters is $0.0127 \pm 0.0063\,\mathrm{Pa}$.
	Nevertheless, in terms of the correction factor this is large or $f_C = 5.25$ and this single data point extends the ground based measurements by a factor of about $2$ (see max. value of $0.025\,\mathrm{Pa}$ in Fig. \ref{fig.shear_stress(pressure)}).
	Normalized by $f_C$ Fig. \ref{fig.shear_stress(gravity)} shows the gravity dependence of the shear stress.
	
	We included the low gravity measurements by \citet{Demirci2019} which used the same particle sample. The data show that 
	\begin{itemize}
		\item the linear dependence on gravity holds or -- noting that the number of data points is not large -- is in agreement to Eq. \ref{eq:tau_erosion_final}.
		\item this agreement requires a correction factor $f_C$. If that was not applied the low gravity data point would not be in agreement to the earlier data points at higher pressure and gravity.
		\item the effective surface energy is in the range of $3\times 10^{-5}$ to $1.1\times10^{-4}\,\mathrm{N}\,\mathrm{m}^{-1}$.
		At ground based experiments gravity dominates over surface energy and the fit to Eq. \ref{eq:tau_erosion_final} is poor. However, close to $0\,g_\mathrm{earth}$, cohesion dominates and makes this quantity more accurate. It shows that cohesion is really low. At zero gravity the shear stress required to initiate erosion is only $0.0011\,\mathrm{Pa}$.
	\end{itemize}
	It has to be noted that the shear stress is lower than the tensile strength assumed for planetesimals e.g. by \citet{Skorov2012}. They did not consider gravity and determined the tensile strength to $0.96\,\mathrm{Pa}$. Obviously, while both quantities come with the same unit, there is a significant difference between threshold shear stress and tensile strength with the latter being higher while the former is responsible for erosion. However, the low value could only be confirmed for the first time in the microgravity experiments.
	
	\section{Astrophysical and Planetary Application}
	\label{sec:astrophysicalapplication}
	
	\subsection{Wind erosion of pebble pile planetesimals}
	
	We apply the model, presented in this work, for wind erosion in slip flow on pebble pile planetesimals. A pebble pile planetesimal which is moving through the gas of the protoplanetary disc with a relative velocity $U_\mathrm{rel}=50\,\mathrm{m}\,\mathrm{s}^{-1}$ might be unstable against wind erosion, if the gas density is high enough. A planetesimal becomes unstable if the wall shear stress $\tau_\mathrm{wall}$ acting on a planetesimal is greater than the threshold shear stress for erosion (see Eq. \ref{eq:tau_erosion}). The wall shear stress acting on the planetesimal depends on the gas density $\rho$ and the friction velocity $u^*$ of the flow around the planetesimal. We use a wind tunnel calibration by \citet{Greeley1980} to obtain the friction velocity $u^*=0.702 \,U_\infty \left(\log_{10} \frac{\rho U_\infty 2R}{\eta} \right)^{-1.59} $ in dependence of the free flow velocity $U_\infty =U_\mathrm{rel}$, planetesimal radius $R$ and gas density $\rho$
	\begin{equation}
		\tau_\mathrm{wall}=\rho u^{*2}= \rho \left[0.702 \, U_\infty \left(\log_{10} \frac{\rho U_\infty 2R}{\eta} \right)^{-1.59}\right]^2.
	\end{equation}
	
	We calculated the ratio $\tau_\mathrm{wall}/ \tau_\mathrm{erosion}$ in dependence of the radial distance $a$ of the planetesimal to the central star and the planetesimal radius $R$. The gravitational acceleration on the surface of a small planetesimal is assumed to be
	\begin{equation}
		g=\frac{4}{3} \pi G \rho_\mathrm{pla} R
	\end{equation}
	with a constant planetesimal bulk density $\rho_\mathrm{pla}=538\,\mathrm{kg}\,\mathrm{m}^{-3}$ \citep{Patzold2019} and the gravitational constant $G$. For the protoplanetary disc we apply the model of the Minimum Mass Solar Nebula, where the radial distance dependence of the gas density is described by \citep{Hayashi1981}
	\begin{equation}
		\rho=\rho_0 \left(\frac{a}{a_0} \right)^{-\frac{11}{4}},
	\end{equation}
	with $\rho_0=1.4\times 10^{-6}\,\mathrm{kg}\,\mathrm{m}^{-3}$ and $a_0=1\,\mathrm{AU}$.\\
	
\begin{figure}
		\centering
		\includegraphics[width=\columnwidth]{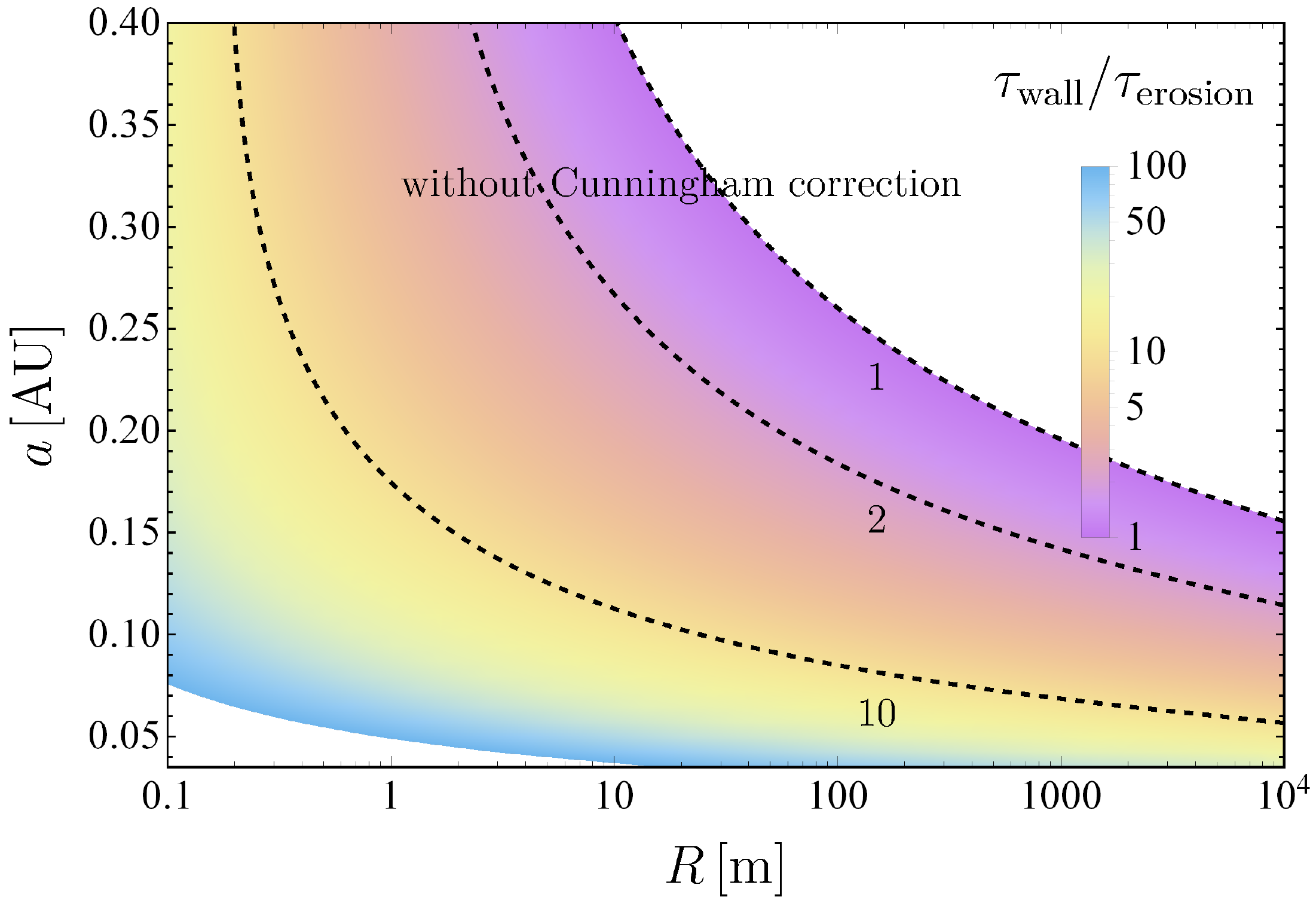}
		\includegraphics[width=\columnwidth]{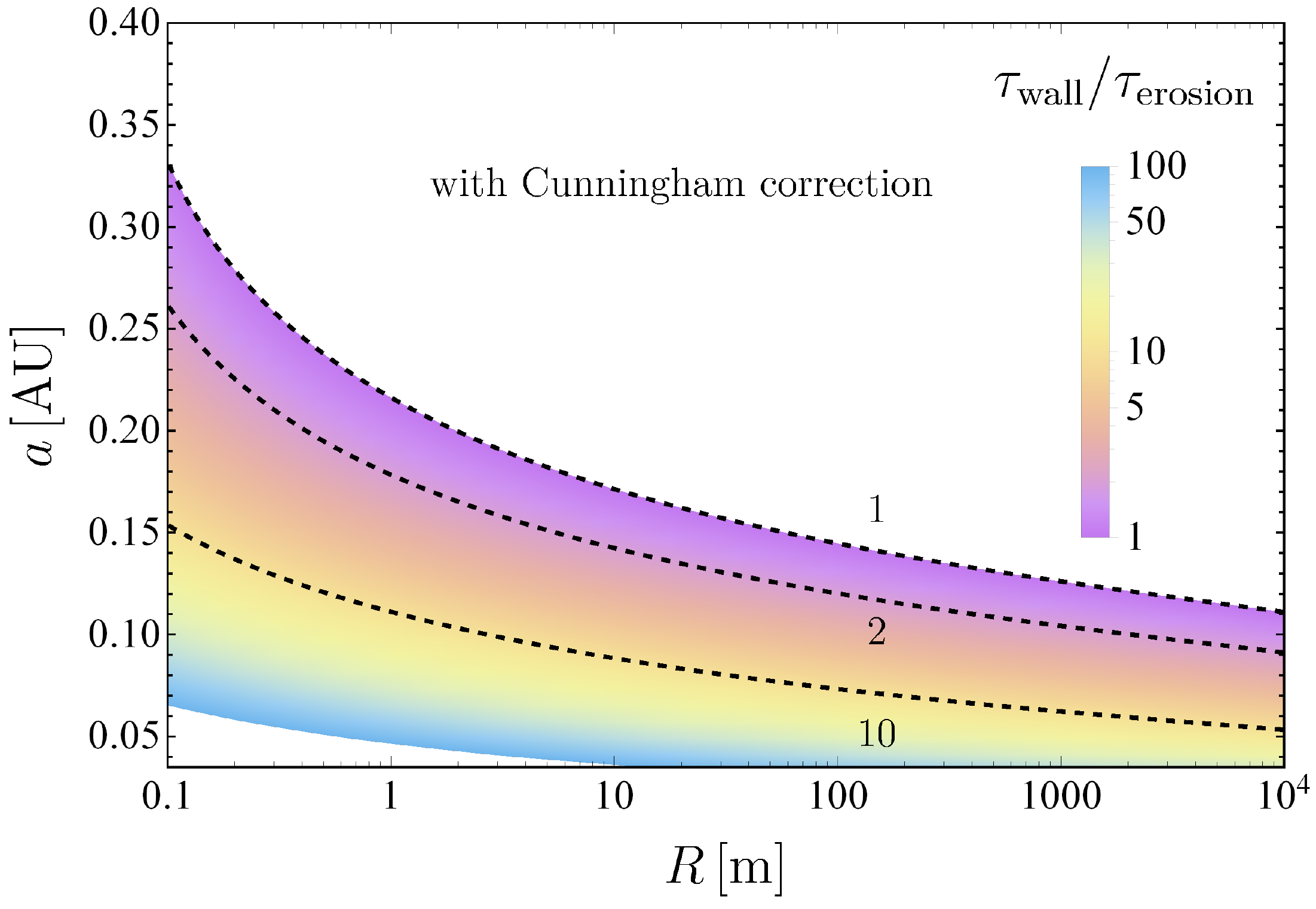}
		\caption{\label{fig.stability_region} $\tau_\mathrm{wall}/ \tau_\mathrm{erosion}$ in dependence of the planetesimal size $R$ and its semi-major axis $a$ for $U_\mathrm{rel}=50\,\mathrm{m}\,\mathrm{s}^{-1}$. The planetesimal density is assumed to be $\rho_\mathrm{pla}=538\,\mathrm{kg}\,\mathrm{m}^{-3}$ \citep{Patzold2019} and the pebble size is assumed to be $d=1\,\mathrm{mm}$. For $\tau_\mathrm{wall}/ \tau_\mathrm{erosion}>1$ erosion occurs. Contour lines for $\tau_\mathrm{wall}/ \tau_\mathrm{erosion}=1, 2, 10$ are included. The lower image is calculated in contrast to the upper one with the Cunningham correction. With Cunningham correction only small planetesimals near to the central star are unstable against erosion by wind.}
	\end{figure}

	In Fig. \ref{fig.stability_region} the ratio $\tau_\mathrm{wall}/ \tau_\mathrm{erosion}$ is plotted in dependence of the radial distance to the central star $a$ and the planetesimal radius $R$. For $\tau_\mathrm{wall}/ \tau_\mathrm{erosion}>1$ the planetesimal will be destroyed by the protoplanetary wind. In this figure we compare the former erosion model by \citet{Shao2000} (Fig. \ref{fig.stability_region}, top) with the new erosion model with Cunningham correction for slip flow (Fig. \ref{fig.stability_region}, bottom).\\
	
	The region in the protoplanetary disc where a planetesimal becomes unstable by wind erosion is much smaller, if slip flow is considered. E.g. a planetesimal with $R=100\,\mathrm{m}$ is eroded at radial distances $a<0.25\,\mathrm{AU}$ without Cunningham correction and $a<0.15\,\mathrm{AU}$ with Cunningham correction (see Fig. \ref{fig.stability_region}). A decimetre sized protoplanetary object is only eroded at radial distances $a<0.35\,\mathrm{AU}$. \citet{Rozner2019} discuss that such objects, which are stable in our threshold based wind erosion, might still be eroded due to gas drag. They discuss this in the context of ablation which might destroy such bodies on much longer timescales, e.g. $\sim 10^{-1}\,\rm Myr$ at $1\,\mathrm{AU}$ \citep{Rozner2019}. We cannot quantify such long term behaviour but only refer to rather fast wind driven erosion here. With that fast destruction in mind, large pebble pile planetesimals with $R=10^4\,\mathrm{m}$ are only eroded at very small radial distances $a<0.1\,\mathrm{AU}$. The region, where planetesimals on circular orbits are destroyed by protoplanetary winds, is smaller than previously extrapolated by \citet{Demirci2019}.
	
	\subsection{Interaction of planetesimals with atmospheres of forming giant planets}
	
	In the core accretion model, a forming giant planet's core gains mass by impacting solids at the same time as his atmosphere grows  \citep{Pollack1996}. If a large solid object impacts it might encounter different fates. It might either just cross the atmosphere and pass the core, be captured by the atmosphere, reach the core and impact or be dismantled one way or the other during his passage of the atmosphere \citep{Brouwers2018, Valletta2019}. However, erosion of pebble pile planetesimals might be another option under certain conditions, especially compared to ablation which might only occur at friction far beyond the threshold shear stress for erosion. Depending on the atmosphere's density, the impact angle, or the impact velocity the planetesimal will leave pebbles within the atmosphere. Depending on the pebble sizes this will change the optical parameters of the atmosphere, which then has a direct impact on the possible gas accretion rates. This is a topic of its own and details of erosion by protoplanetary atmospheres are certainly beyond the scope of this paper but it might change the formation of giant planets significantly.
	
	\subsection{Saltation on planetary surfaces}
	
	\textbf{Mars:}
	Our results are not only important for pre-planetary phases but also for planetary surfaces with low pressure atmospheres. Certainly, the best example is Mars, where eolian motion of dunes has been observed at certain locations and times in an atmosphere of only $6\,\rm mbar$ on average \citep{Baker2018}. Particles of about $100\,\rm \mu m$ in size are frequently considered as the particles that move best \citep{Greeley1980}. Our experiments show for the first time that the flow at these conditions is already well in the transition regime. That means that the required friction velocities ($+7\,\%$) or threshold shear stresses ($+15\,\%$) are higher than what might be scaled from higher pressure conditions. This does not ease the problem of lifting grains on Mars, though the results by \citet{Musiolik2018} indicate that conditions can be sufficient for saltation. Why dust is moved at only $1\,\rm mbar$ on Arsia Mons gets ever more enigmatic though \citep{Reiss2009}. Anyhow, if extrapolated from laboratory experiments at higher pressure to Martian conditions, a potentially strong increase in required shear stress to initiate saltation should be kept in mind.\\
	\noindent
	\textbf{Pluto:}
	With $0.06\,g_\mathrm{earth}$ Pluto roughly matches the $g$-levels used during the parabolic flights. While the atmosphere is tenuous with approximately $3\times 10^{-3}\,\mathrm{mbar}$ ambient pressure, this was tested during the parabolic flights for erosion. Dunes have been observed on Pluto consisting of methane-ice \citep{Telfer2018}. Therefore, wind has to be active there. We could not see any erosion though at the highest shear stresses we could apply. Estimating from the equation found above, we would expect at least $0.042\,\mathrm{Pa}$ as shear stress for a grain with $d=425\,\mu\mathrm{m}$ or in terms of threshold friction velocity $u^*_\mathrm{erosion}=42.6\,\mathrm{m}\,\mathrm{s}^{-1}$. With a logarithmic wind profile $u=u^*_\mathrm{erosion}/\kappa\log\left( h/h_0 \right)$ with $\kappa=0.4$ as von Karman constant and $h_0 \approx d/30$ the surface roughness, this results in a wind speed of $622\,\mathrm{m}\,\mathrm{s}^{-1}$ in a height of $h=10\,\mathrm{m}$. This is much more than the wind speeds predicted on Pluto \citep{Zalucha2016} and therefore it is not likely that the onset of saltation is provided by wind.
	
	\section{Conclusions}
	\label{sec:conclusion}
	We could show in laboratory experiments that the gas drag on surface grains is determined by the grain's Knudsen number corrected by a factor $\beta = 0.67$. This means that at Knudsen numbers approaching 1 or larger values, the drag force has to be corrected and results in
	higher threshold shear stress for erosion. The equation describing the shear stress is then Eq. \ref{eq:tau_erosion_final}.
	
	This will already be important at Martian pressure, which at high elevations can be as low as $1\, \rm mbar$. This has not been considered before, is certainly of importance but details are beyond the scope of this paper.
	
	We had erosion of pebble planetesimals in protoplanetary discs in mind. This is strongly influenced by this low pressure correction. Uncorrected, a shear stress of only $4.8 \times 10^{-4}\, \rm Pa$ is needed to start destroying a planetesimal with a size of $R=10\, \rm km$ at $1 \, \rm AU$. This is very low and assumes that the surface energy or sticking force of our sub-mm grains are similar to the sticking forces of dust pebbles.
	Corrected, the shear stress at $1\,\rm AU$ is $1.8 \times 10^{-1}\, \rm Pa$ in the Minimum Mass Solar Nebula which is a factor 380 larger. Outwards of $1\,\rm AU$ the correction factor increases. It decreases further inwards. 
	In spite of the correction, planetesimals are still prone to destruction in the inner protoplanetary disc on circular orbits and to larger distances if the orbit is only slightly eccentric.
	
	\section*{Acknowledgements}
	
	This project is funded by DLR space administration with funds provided by the BMWi under grant 50 WM 1760. N. S. is funded by the DFG under grant WU 321/16-1. T. B. is funded by the DFG under grant WU 321/18-1. We appreciate the constructive review by the anonymous referee. We also thank M. Aderholz who helped bringing the idea of this experiment to real life.
	
	
	
	
	\bibliographystyle{mnras}
	\bibliography{demirci} 

	
	
	
	%
	%
	

	\bsp	
	\label{lastpage}
\end{document}